\documentclass[%
 reprint,
nofootinbib,
 amsmath,amssymb,
 showkeys,
 aps,
]{revtex4-1}

\usepackage{graphicx}% Include figure files
\usepackage{dcolumn}% Align table columns on decimal point
\usepackage{bm}% bold math
\usepackage{hyperref}% add hypertext capabilities
\begin{document}

\title{On the difficulty of generating gravitational wave turbulence in the early universe}

\author{Katy Clough}
 \email{katy.clough@phys.uni-goettingen.de}
\author{Jens C. Niemeyer}
 \email{jens.niemeyer@phys.uni-goettingen.de}
\affiliation{${}$Institut f{\"u}r Astrophysik, Georg-August Universit{\"a}t, Friedrich-Hund-Platz 1, D-37077 G{\"o}ttingen, Germany}

\begin{abstract}
A recent article by Galtier and Nazarenko \cite{Galtier:2017mve} proposed that weakly nonlinear gravitational waves could result in a turbulent cascade, with energy flowing from high to low frequency modes or vice versa. This is an interesting proposition for early universe cosmology because it could suggest some ``natural'' initial conditions for the gravitational background. In this paper we use the ADM formalism to show that, given some simple and, arguably, natural assumptions, such initial conditions lead to expansion (or collapse) of the spacetime on a timescale much faster than that of the turbulent cascade, meaning that the cascade is unlikely to have sufficient time to develop under general conditions. We suggest possible ways in which the expansion could be mitigated to give the cascade time to develop.
\end{abstract}

%\keywords{gravitational waves, turbulence, early universe cosmology}
\maketitle

\section{\label{sec:intro}Introduction}

Cosmological inflation provides a widely accepted framework for the earliest moments in the history of our universe \cite{Guth:1980zm,Linde:1981mu,Albrecht:1982wi,Starobinsky:1980te}. Stretching spacetime perturbations of the order of the initial horizon size to scales far outside our present particle horizon is both its biggest strength -- by establishing a long-lived homogeneous region of space -- and weakness, shielding us from observable information about its prior state. Exploring the initial conditions for inflation from a theoretical perspective, while speculative by nature, is nevertheless important for addressing questions of ``naturalness'' and extended theories of the early universe such as bouncing cosmologies or false-vacuum tunnelling in landscape scenarios \cite{Steinhardt:2001st, Khoury:2001bz, Khoury:2001wf, Coleman:1977py, Callan:1977pt}. 

Without a specific scenario in mind (which normally also requires an initial state), the search for a ``natural'' state of spacetime can be guided by attractor solutions with universal properties. As an alternative to highly symmetric vacuum or thermal states, non-equilibrium physics provides classes of near-universality such as hydrodynamical turbulence for strong nonlinearities \cite{frisch:1995} or ``wave turbulence'' for weakly nonlinear interactions in a wide variety of theories \cite{NazarenkoBook}. In both cases, nonlinearities give rise to cascades that establish power-law spectra across a range of wavenumbers that are sufficiently separated from the largest and smallest scales of the problem. Clearly, finding such a state for spacetime, driven by some unspecified mechanism at earlier times or larger scales, would be an attractive starting point for further investigations in the context of inflationary initial conditions.

A recent article by Galtier and Nazarenko \cite{Galtier:2017mve} proposed that weakly nonlinear gravitational waves could result in an energy cascade obeying the dynamics of wave turbulence.  Further details can be found in \cite{NazarenkoBook, KolmogorovBook}. 
Whilst propagating gravitational waves from astrophysical events are small and thus very well described by the linear approximation, in the early universe, for example, at a time prior to inflation, a gravitational wave background could be large enough to result in nonlinear self-interactions, and one might expect that it would contain a random spectrum of frequencies. This proposition is particularly interesting because it predicts an inverse cascade of waves towards the lowest accessible wavenumbers analogous to Bose-Einstein condensation, suggesting a potential partial solution to the initial conditions problem for inflation. It might also provide a new basis for numerical studies of this problem \cite{East:2015ggf, Clough:2016ymm, Clough:2017efm}.

In this paper, we identify a serious problem of this scenario related to the backreaction of gravitational waves on the expansion of their background spacetime. The energy density of a gravitational wave with (small) amplitude $h$ and frequency $\omega$ is of order $m_{\rm pl}^2 \omega^2 h^2$, resulting in Hubble expansion or contraction with rate $H \sim \tau_{\rm exp}^{-1} \sim \omega h$. On the other hand, \cite{Galtier:2017mve} show that the typical timescale for the gravitational wave cascade is given by $\tau_{\rm cas} \sim \omega^{-1} \epsilon^{-4}$, where $\epsilon \sim h$ is a small parameter that characterises the strength of the nonlinearity. We therefore argue that under very general conditions, the timescale required to establish the gravitational wave cascade is longer than the expansion time estimated from backreaction by a factor of order $h^{-3}$. Any nonlinear cascade would therefore be quenched before a universal power-law spectrum or a significant low-wavenumber condensate could be produced.

The paper is structured as follows. In section \ref{sec:ICs} we discuss initial conditions for turbulent gravitational waves, using the ADM formalism of general relativity. In particular we focus on the way in which such initial conditions must satisfy the Hamiltonian and momentum constraints, subject to assumptions about the initial degrees of freedom. Similar arguments could be made simply by assuming an approximately homogeneous FRW universe, but the analysis here is more generally applicable. In section \ref{sec:timescales} we discuss the timescales involved, and explain why the initial conditions described would generally lead to a collapse or expansion of the spacetime on a timescale much faster than that of the turbulent cascade. We conclude in section \ref{sec:conclusions} and propose ways in which the problem could be mitigated. 

\section{\label{sec:ICs}Initial conditions}

In the ADM decomposition \cite{PhysRev.116.1322}, the 4 dimensional spacetime metric is decomposed into a spatial metric on a 3 dimensional spatial hypersurface, $\gamma_{ij}$, and an extrinsic curvature $K_{ij}$, which both evolve along a local time coordinate $t$ in a chosen gauge.

We will first briefly describe the ADM formalism and conformal decomposition in order to set the conventions used. We will then discuss how the degrees of freedom are set, partly by well motivated choices, and partly subject to physical constraints.

\subsection{ADM decomposition and notation conventions}

The line element of the ADM decomposition is
\begin{equation}
ds^2=-\alpha^2\,dt^2+\gamma_{ij}(dx^i + \beta^i\,dt)(dx^j + \beta^j\,dt)\,,
\end{equation}
where $\gamma_{ij}$ is the metric of the initial space-like hypersurface.

The extrinsic curvature is related to the ``time derivative'' of the spatial metric, defined by
\begin{equation}
K_{\mu \nu} = - 1/2 \pounds_{\vec{n}} \gamma_{\mu \nu} ~ \label{eqn:Kabdefinition}
\end{equation}
where $\vec{n}$ is the normal vector to the spatial hypersurface.

Initial conditions consist in specifying the metric $\gamma_{ij}$ and the extrinsic curvature $K_{ij}$ everywhere on some initial spatial slice. The values are subject to the Hamiltonian and Momentum constraints imposed by the Einstein equations, which in the absence of matter are
\begin{equation}
\mathcal{H} = R + K^2-K_{ij}K^{ij} = 0 \label{eqn:HamCon1}
\end{equation}
\begin{equation}
\mathcal{M}_i = D^j (\gamma_{ij} K - K_{ij}) = 0 ~ . \label{eqn:MomCon1}
\end{equation}

For the purposes of choosing initial conditions, the induced metric on the 3D hyperslice is usually decomposed into a scalar conformal factor $\psi$ and a conformal metric $\bar \gamma_{ij}$ with
\begin{equation}
\gamma_{ij}=\psi^4 ~ \bar\gamma_{ij} ~ , \quad \det\bar\gamma_{ij}=1 ~ .
\end{equation}
The extrinsic curvature is decomposed into its trace, $K=\gamma^{ij}\,K_{ij}$, and its trace-free part \begin{equation}
A_{ij} = K_{ij} - \frac{1}{3}\gamma_{ij} K ~.
\end{equation}
It is also useful to define the ``natural'' conformal rescaling of this traceless part, such that $\bar A_{ij} = \psi^{2} {A}_{ij}$.

\subsection{Setting the metric degrees of freedom}

We have some freedom to specify the initial conditions on the metric and extrinsic curvature, corresponding partly to the freedom to specify coordinates on the initial slice, and partly to setting dynamical degrees of freedom. In this section we list our choices and derive the resulting constraints on the remaining degrees of freedom.

A key assumption we will make throughout is that of statistical isotropy and on average zero intrinsic curvature of the spatial slices. This results in a periodic domain with some repeating ``typical patch'' in the overall spacetime, such that the patch of space under consideration is topologically a torus. We also assume that no matter is present, such that we consider only vacuum solutions of the Einstein equation.

We will discuss the effect of varying the assumptions in our conclusions, in section \ref{sec:conclusions}.

\subsubsection{The Momentum Constraint}

With the conformal decomposition described above, and in the absence of matter, the momentum constraint Eqn. (\ref{eqn:MomCon1}) can be written as
\begin{equation}
\partial_j \bar{A}^{ij}  -\frac{2}{3}\psi^6 \partial^j K = 0. \label{eqn:MomCon2}
\end{equation}
where we make the simplifying assumption that the metric is conformally flat everywhere on the initial slice, that is
\begin{equation}
\tilde{\gamma}_{ij} = \delta_{ij} ~ .  \label{eqn:conformalflatmetric}
\end{equation}
This seems a reasonable assumption in a spacetime with no overall preferred spatial direction, ie, with gravitational waves propagating in all directions equally, and on average zero intrinsic curvature. It also implies that we have chosen a moment at which the gravitational wave perturbations are instantaneously zero (their time derivative is non zero, and set via $A_{ij}$ below).

We also choose the trace of the extrinsic curvature $K$ to be a spatial constant. This is equivalent to setting a constant rate of expansion or contraction across the spatial slice. (In the homogeneous and isotropic FRW metric, $K$ is related to the Hubble constant as $K=-3H$.)

Given this choice, the momentum constraint reduces to the requirement that $\bar A_{ij}$ is transverse\footnote{The longitudinal part can be shown to be zero for a vacuum spacetime with periodic boundary conditions (see, for example, \cite{Clough:2017efm}).}
\begin{equation}
\partial_j \bar{A}^{ij} =0. \label{eqn:MomCon3}
\end{equation}

Since we are working on a torus of length $L$, we can Fourier decompose $\bar A^{ij}$, as
\begin{equation}
\bar A^{ij}=\sum_{n_1,n_2,n_3}c_{(n_1,n_2,n_3)}^{ij}\exp\left[\sum_k \frac{2\pi i n_k x^k}{L}\right]\,.
\end{equation}
The conditions that $\bar{A}^{ij}$ be transverse and traceless imply
\begin{equation}
\sum_i n_i c_{(n_1,n_2,n_3)}^{ij}=0\quad\text{and}\quad\sum_i c_{(n_1,n_2,n_3)}^{ii}=0\,.\\
\end{equation}
We see that these equations impose four conditions on six independent matrix elements, leaving us with two degrees of freedom. In the linear regime, these correspond to the two polarization states of the graviton.

For approximate ``gravitational waves", the amplitude of each mode will be related to its wavenumber $\bf{n}$, by
\begin{equation}
c^{ij}_{\bf{n}} \sim \omega ~ h \label{eqn:A_ij_modes},
\end{equation}
where $h$ is the amplitude of the gravitational waves and $\omega = 2\pi |{\bf{n}}|/L$ a value which sets the frequency of the fluctuations relative to the scale of statistical isotropy $L$.

Since the momentum constraint is linear, we can in principle superpose any number of modes to construct an $\bar{A}_{ij}$ with the desired power spectrum of gravitational wave frequencies for investigation of the turbulent cascade.

\subsubsection{The Hamiltonian Constraint}

Having set the values for $\bar A^{ij}$, we must now solve the Hamiltonian constraint, which again can be written in terms of the decomposed variables as
\begin{equation}
    \partial_i\partial^i\psi - \frac{\psi \bar{R}}{8} + \frac{\psi^{5}}{8} \left({A}_{ij} {A}^{ij} - \frac{2}{3} K^2\right) = 0 \label{eqn:HamCon2}\,.
\end{equation}

Given that the conformal metric is flat, $\bar{R}$ is zero, and the $\partial_i\partial^i \psi$ is the (flat space) Laplacian of the conformal factor.

We already chose the trace of the extrinsic curvature $K$ to be a spatial constant, and we might ideally like to set it to be small, to prevent any expansion of the spacetime. However, its value is imposed by the requirement to satisfy the Hamiltonian constraint with smooth periodic boundary conditions (imposed by the compactification of the space on the isotropic scale), since the Laplacian of the conformal factor must average to zero over the volume. Thus we require
\begin{equation}
K = \pm \sqrt{\langle 3/2 ~ A_{ij} A^{ij}}  \rangle  \,, \label{eqn:Kavg}
\end{equation}
where a negative value means that the spacetime is expanding, and a positive one that it is contracting, and $\langle X \rangle = {\cal V}^{-1} \int X~d{\cal V}$ indicates the average over the spatial volume ${\cal V}$ of the quantity $X$.

Note that in principle we do not know the values of $\psi$, and therefore $A_{ij} = \psi^2 \bar A_{ij}$ which are required for the calculation of $K$ in equation (\ref{eqn:Kavg}). However, we can assume that
\begin{equation}
K \approx \pm \sqrt{\langle 3/2 ~ \bar A_{ij} \bar A^{ij}}  \rangle  \,, \label{eqn:Kavg2}
\end{equation}
since even for fairly large gravitational wave fluctuations, $\psi\sim 1$. For a single mode, equation (\ref{eqn:A_ij_modes}) then implies that
\begin{equation}
K \sim \omega ~ h. \label{eqn:Kavg3}
\end{equation}

\section{\label{sec:timescales}Timescale of expansion or collapse}

We see that with the choices made it is not possible to set $K=0$ on the initial slice. We have not made any strong assumptions about the scale of the gravitational wave perturbations (their amplitude or frequency), but in the limit where they have a small amplitude and high frequency relative to the spatial scale L, the requirement to have a non zero $K$ can be related to the Isaacson stress energy tensor \cite{Isaacson:1967zz}
\begin{equation}
    T_{ab} = \frac{1}{32\pi} \partial_a h^{TT}_{cd} \partial_b h_{TT}^{cd} ~.
\end{equation}
The $T_{00}$ component of this tensor can be identified with the term $A_{ij}A^{ij}$ in the Hamiltonian constraint, and serves as an additional energy density, creating an expanding (or contracting) spacetime.

We can ask how much this expansion will affect the turbulent cascade. If the timescale of expansion is much longer than that of the turbulent cascade, it can safely be neglected. The timescale of the expansion $\tau_{exp}$ is of order $1/K$, and thus for gravitational waves of a single frequency $\omega$ and amplitude $h$
\begin{equation}
    \tau_{exp} \sim \frac{1}{h \omega}
\end{equation}
where $\omega \sim 1/\ell$ where $\ell < L$ is the length scale of the gravitational waves.

In \cite{Galtier:2017mve} it is found that the timescale for the turbulent cascade is of order
\begin{equation}
    \tau_{cas} \sim \frac{\tau_{NL}^4}{\tau_{GW}^3}
\end{equation}
where
$\tau_{NL} \sim \ell/h$ and $\tau_{GW} \sim 1/\omega$. Thus, the ratio of the timescales is
\begin{equation}
    \frac{\tau_{cas}}{\tau_{exp}} \sim \frac{1}{h^3}
\end{equation}
and since for weakly interacting waves $h \ll 1$ the timescale for the cascade is much longer than that for the expansion or contraction of spacetime.

We have considered only a single frequency, but adding additional modes of gravitational waves will only decrease $\tau_{exp}$ relative to this calculation. It is therefore probable that for a spectrum of modes distributed about some mean $\ell$, the gravitational wave content will be diluted by the expansion (or that the spacetime will collapse) before it can undergo the turbulent cascade.

\section{\label{sec:conclusions}Conclusions}

We have shown that simple choices of initial conditions lead to a case in which the timescale of the cascade would be far longer than the timescale of the expansion or contraction of the spacetime in which they propagate, thus it is unlikely that the cascade would have time to occur.

To summarise, the choices which led to this conclusion are:
\begin{enumerate}
    \item Statistical isotropy, and on average zero intrinsic curvature of the spatial slices, such that the spatial patch is topologically a torus
    \item No matter and zero cosmological constant
    \item An initially spatially homogeneous conformally flat metric
    \item The trace of the extrinsic curvature is a constant on the slice
\end{enumerate}
Of these choices, the assumption of statistical isotropy and on average flatness is the main driver of the expansionary behaviour.

Relaxing the assumption of on average flatness could alleviate the expansionary problem if one imposes a topology for which it is possible to have non negative $^{(3)}R$ ($S^3$ and $S^2 \times S^1$, see \cite{Barrow:1985}).

As an alternative to imposing such topologies, one could relax the assumption of statistical isotropy and investigate the case of asymptotic flatness. The waves would then be concentrated in a localised patch, which would generate a gravitational well within the otherwise flat spacetime. Whilst this is a possibility, it does not correspond to the quasi-uniform covering of the Universe which is proposed in \cite{Galtier:2017mve}.

Matter with a positive energy density only results in a larger expansion or contraction rate, and thus only exacerbates the timescale problem. However, one could consider adding a negative cosmological constant of the right magnitude in order to counteract the expansion, to result in a net stationary spacetime. 

The last two conditions have minimal physical effect and were imposed mainly to simplify the calculation. Relaxing the condition of a constant $K$ would still require that on average, $K^2 - A_{ij}A^{ij} = 0$, and thus would still give us regions of non zero $K$ corresponding to regions where gravitational wave densities are high. Equally allowing the metric to locally deviate from conformal flatness, (but without changing the average intrinsic curvature, so the topology is still toroidal), would be unlikely to change the overall expansionary behaviour.

\section{Acknowledgements}
We thank the authors of \cite{Galtier:2017mve} for helpful discussions regarding their work. KC thanks Eugene Lim and Raphael Flauger for their input to the work in \cite{Clough:2017efm}, in which several of the ideas underlying this article were developed.

\bibliography{mybib.bib}

\end{document}